\documentclass[12pt,latexsym,multicol,amsfonts,epsf,xy,amsbold,amsmath,amssymb,m
athsym]{article}

\usepackage[frame,line,arrow,matrix,tips]{xy}
\input{Qcircuit}

\def\C{{\mathbf{C}}}
\def\N{{\mathbf{N}}}
\def\R{{\mathbf{R}}}

\date{October 27, 2006}

\newcommand{\cL}{{\cal L}}

\newcommand{\beq}{\begin{equation}}
\newcommand{\eeq}{\end{equation}}
\newcommand{\beqy}{\begin{eqnarray}}
\newcommand{\eeqy}{\end{eqnarray}}

\newcommand{\polylog}{{\rm polylog}}
\newcommand{\poly}{{\rm poly}}

\newtheorem{Definition}{Definition}
\newtheorem{Lemma}{Lemma}
\newtheorem{Theorem}{Theorem}

\newenvironment{Definition*}{{\bf Definition}}{}

\title{BQP-complete Problems Concerning  Mixing Properties 
of Classical Random Walks on Sparse Graphs}

\author{Dominik Janzing\thanks{e-mail: janzing@ira.uka.de}\\ 
\small Institut f{\"u}r Algorithmen und Kognitive Systeme,
Universit{\"a}t Karlsruhe,\\[-1ex] \small Am Fasanengarten 5,
D-76\,131 Karlsruhe, Germany\\ 
\\
Pawel Wocjan\thanks{e-mail: wocjan@cs.ucf.edu}\\
\small School of Electrical Engineering and Computer Science\\[-1ex]
\small University of Central Florida\\[-1ex]
\small Orlando, FL 32816, USA}

\begin{document}

\maketitle

\begin{abstract}
We describe two BQP-complete problems concerning properties of sparse 
graphs having a certain symmetry.  The graphs are specified by 
efficiently computable functions which output the adjacent vertices for 
each vertex. Let $i$ and $j$ be two given vertices.  The first problem consists in estimating the difference between the number of paths of length $m$ from $j$ to $j$ and those which from $i$ to $j$, where $m$ is polylogarithmic in the number of vertices.  The scale of the estimation accuracy is specified by some a priori known upper bound on the growth of these differences with increasing $m$. The problem remains BQP-hard for regular graphs with degree $4$.

The second problem is related to continuous-time classical random walks.  The walk starts at some vertex $j$. The promise is that the difference of the probabilities of being at $j$ and at $i$, respectively, decays with $O(\exp(-\mu t))$ for some $\mu>0$.  The problem is to decide whether this difference is greater than $a\exp(-\mu T)$ or smaller than $b\exp(-\mu T)$ after some time instant $T$, where $T$ is polylogarithmic and the difference $a-b$ is inverse polylogarithmic in the number of vertices.  Since the probabilities differ only by an exponentially small amount, an exponential number of trials would be necessary if one tried to answer this question by running the walk itself.

A modification of this problem, asking whether there exists a pair of nodes
for which the probability difference is at least $a\exp(-\mu T)$, is QCMA-complete. 
\end{abstract}

\section{Introduction}
Although it is commonly believed that quantum computers would enable us to solve several mathematical problems more efficiently than classical computers, it is difficult to characterize the class of problems for which this is expected to be the case.  The class of 
problems which can be solved efficiently on a classical
computer by a probabilistic algorithm is the complexity class BPP.  The quantum counterpart of this complexity class is BQP, the class of problems that can be solved efficiently on a quantum computer with bounded error.  Thus, the exact difference between the complexity classes BPP and BQP remains to be understood.

An important way of understanding a complexity class is to find problems which are complete for the latter.
Meanwhile, some examples of BQP-complete problems are known \cite{KnillQuadr,PawelYard,aharonov-2006-,WZ:06,DiagonalEntry}. 
For the results presented here, the ideas of \cite{DiagonalEntry} are crucial.  We rephrase the main idea and proof of this article since they 
provide the basis for the present work.  The problem is stated in terms of sparse matrices. Following  \cite{ATS,ChildsDiss,BACS:06}, we call  
an $N\times N$ matrix $A$ sparse if it has no more than $s=\polylog(N)$ non-zero entries in each row and there is an efficiently computable function  which specifies for a
given row the non-zero entries and their positions.  

We have formulated a decision problem concerning the estimation of the $j$th diagonal entry of the $m$th power of $A$ and proved that it is BQP-complete.  The proof that this problem is BQP-hard as well as the proof that it is in BQP both rely on the following observation.  The $j$th diagonal entry of the $m$th power is the $m$th statistical moment of the probability distribution of outcomes when a measurement of the ``observable'' $A$ is applied to the $j$th basis vector. Thus, the BQP-complete problem consists in estimating the probabilities of the outcomes in an $A$-measurement. Moreover, 
we have shown in \cite{DiagonalEntry} that the problem remains BQP-hard
if it is restricted to matrices with entries $\pm 1,0$. 

It is not surprising that questions related to decompositions of vectors with respect to eigenspaces of large matrices lead to hard computational problems and that quantum computer could be more powerful in dealing with such problems. After all, the dynamics of a quantum system is determined by the spectral resolution of its initial state.  In this context, the interesting questions and challenges are to determine to what extent it is possible to identify 
{\it natural} problems related to spectral resolutions which are not mere reformulations of questions about the dynamical behavior of quantum systems. 

Calculating spectral decompositions with respect to large real symmetric matrices is certainly not a problem
which only occurs in quantum theory.  For example, methods in data analysis, machine learning, and signal processing rely strongly on the efficient solution of the principle component analysis \cite{Schoelkopf}. 
However, data matrices typically occurring in these applications are in general not sparse and, even worse, their entries are usually not specified by an efficiently computable function. The data are rather given by observations.

In our opinion, it is an interesting challenge to construct BQP-problems concerning spectra of sparse matrices having only $0$ and $1$ as entries since problems of this kind are related to combinatorial problems in graph theory.  It is quite natural to assume that a graph is sparse in the above sense;
the efficiently
computable function specifying the non-zero entries in each row is then a function which describes the set of neighbors of a given vertex and
it is assumed that the number of neighbors is only polylogarithmic in 
the number of nodes.
 
The paper is organized as follows.  
In Section~\ref{Lipschitz} we present a quantum algorithm for measuring functions of observables.  This algorithm uses the quantum phase estimation algorithm as a subroutine.  We use this algorithm to show that 
the problems ``Diagonal Entry Estimation'' (Section~\ref{Sec:DiagonalEntry}), ``Difference of Number of Paths'' (Section~\ref{Paths}), and ``Decay of Probability Differences'' (Section~\ref{Mixing}) are all in BQP. 
Even though the first problem was already described in \cite{DiagonalEntry}
we have first to extend these results
in Section~\ref{Sec:DiagonalEntry} 
since the proofs for the BQP-completeness of the 
other two problems are based on such an extension.  
In Section~\ref{Paths} we present a BQP-complete problem which (at first) strongly resembles the problem ``Diagonal Entry Estimation''.  The important 
difference is that it deals with adjacency matrices (that is, $0$-$1$-matrices) instead of matrices with entries $\pm 1,0$.  We describe how the problem is
related to mixing properties of a classical random walk with 
discrete time.
In Section~\ref{Mixing} we describe a BQP-complete problem concerning the mixing properties of continuous-time classical random walks on regular sparse graphs. In Section~\ref{Sec:QCMA} we describe a modification of the mixing problem
which is QCMA-complete.

\section{Measuring continuous functions of observables}\label{Lipschitz}

Before we explain what it means to measure functions of observables we have to introduce some terminology.  The spectral measure induced by a self-adjoint
$N\times N$-matrix  $A$ and a state vector 
$|\psi\rangle \in \C^N$ is defined as the measure on $\R$ supported by
the set of eigenvalues $\lambda_j$ of $A$ with probabilities 
$p_j:=\langle \psi | Q_j|\psi\rangle$, where $Q_j$ is the spectral projection 
corresponding to $\lambda_j$. 

Let $f:\R\rightarrow \R$ be a function.  The spectral measure induced by $f(A)$ and $|\psi\rangle$ is then the measure on $\R$ supported by the values $f(\lambda_j)$ with probabilities $p_j$, where $p_j$ and $\lambda_j$ are as above.

We define measuring a function $f$ of an observable $B$ in the state $|\psi\rangle$ as a quantum process which allows us to sample from a probability distribution that coincides with the spectral
measure induced by $f(B)$ and $|\psi\rangle$.  Note that when $f$ is the identity function, then this corresponds what is considered to be a von-Neumann measurement of an observable in quantum mechanics.  As in \cite{DiagonalEntry} the main tool for implementing such measurements is the quantum phase estimation algorithm \cite{ClevePhase}.  The idea is that for each observable $B$ with $\|B\|<\pi$ we
can implement the measurement of $B$ by applying quantum phase
estimation to the unitary $\exp(iB)$, i.e., the map that describes 
the corresponding dynamics according to the Hamiltonian $-B$. 
For the class of self-adjoint operators considered here an efficient 
simulation of the dynamics 
is indeed possible since they are sparse \cite{ATS,ChildsDiss,BACS:06}.  
By applying the procedure to 
a given state vector $|\psi\rangle$ we can sample from the spectral measure
induced by $B$ and $|\psi\rangle$. 

The main statement of this section is that we can also sample from the
spectral measure induced by $f(B)$ since we  
can implement such von-Neumann
measurements of $B$ and then apply $f$ to the outcomes. 
Hence, no implementation of $\exp(if(B))$ is needed 
(this is important because,  for sparse $B$, the matrix 
$f(B)$ will in general not be sparse
and therefore it is not clear how to simulate the corresponding dynamics).
In particular, we can estimate the expectation value of
a given $f$ with respect to the spectral measure.
However, for our proofs it is essential how the accuracy behaves
when certain smoothness assumptions are posed on $f$.
This is made precise in  Lemma~\ref{FuncLemm}.

But first we have to recall how quantum phase estimation works 
\cite{ClevePhase}.
We add a $p$-qubit ancilla register to the $n$ qubits on which $V:=\exp(iB)$ 
acts. Then we replace the circuit implementing $V$ with 
a controlled-$V$ gate 
by replacing each elementary gate with its controlled analogue
(note that this cannot be done in a black-box setting). 
The ancilla register is initialized
to the equally weighted positive superposition of all binary words,
then we use copies of the controlled gate to 
apply the $2^j$th power of $V$  to the $n$ qubits if the $j$th ancilla qubit is in the 
state $|1\rangle$.  Finally, we apply the inverse Fourier transform to the ancilla register and measure the ancillas in the computational basis.
Given that the $n$-qubit register was in an eigenstate
of $V$ with eigenvalue $\exp(i2\pi \varphi)$ the obtained binary word 
$a\in \{0,1,\dots 2^p-1\}$ 
provides a good estimation for $2\pi \varphi$ in the following sense:
\begin{equation}\label{thetaeta}
\mathrm{Pr}(|\varphi -a /2^p| <\eta )>1-\theta\,,
\end{equation}
where $\theta,\eta >0$ and the number of ancillas is given by \cite{NC}
\begin{equation}\label{pDef}
p:=
\lceil \log(1/\eta)\rceil + \lceil \log\big(2+(1/(2\theta)\big) \rceil\,.
\end{equation}
Here, the distance $|\varphi-a/2^p|$ is to be understood with
respect to the cyclic topology of the unit circle
which identifies $\varphi=1$ with $\varphi=0$. 
In contrast to the convention in \cite{NC} we reinterpret throughout 
the paper $a$ as phases $x$ in the interval $[-\pi,\pi)$ 
by defining $x:=a \,2\pi/2^p$ if $a\leq 2^{p-1}$ and
$x:= a \,2\pi/2^p -2\pi$ otherwise.  We refer
to $x$ as the {\it outcome} of the measurement procedure. 
Then we have:

\begin{Lemma}[Continuous Functional Calculus]\label{FuncLemm}${}$\\
Given a 
self-adjoint $n$-qubit operator $B$ with $\|B\|\leq 1$
and a quantum circuit $U$ with $\|U-\exp(iB)\|\leq \delta$ for
some $\delta >0$ such that the decomposition of $U$ into elementary gates 
is known. 
Let  $|\psi\rangle$ be an $n$-qubit state 
whose decomposition into
$B$-eigenvectors contains only eigenvectors whose eigenvalues are in the 
closed (but possibly infinite) 
interval $I$.  Let $f:I\rightarrow \R$ be a Lipschitz continuous
function with constant $K$, i.e., $|f(x)-f(y)| \leq |x-y| \,K $ for all $x,y\in I$.

First, we apply the phase estimation (with unitary $U$) 
to $|\psi\rangle$, where the number $p$ of 
ancilla qubits is given by eq.~(\ref{pDef}). 
Next, we apply $f$ to the outcome $x$ if $x\in I$, otherwise
we replace $x$ by the value in $I$ closest to it and apply $f$ to the latter.
The expectation value of the random variable 
$f(X)$ in this experiment, 
denoted by $E_{|\psi\rangle}(f(X))$, 
satisfies
\[
|E_{|\psi\rangle} (f(X)) - \langle \psi|f(B)|\psi\rangle| <  
(2\theta  + \delta 2^{p+1})  \, \|f\|_\infty 
+2\pi K \eta\,,
\]
where $\eta,\theta$ and $p$ are related to each other as in eq.~(\ref{pDef})
and $\|f\|_\infty$ denotes the norm given
by the supremum of $|f(x)|$. 
\end{Lemma}

\medskip
\noindent
Proof: 
We first assume we could implement $V:=\exp(iB)$ instead of its
approximation $U$.  In this case, if we apply phase estimation to an eigenvector
$|\psi_j\rangle$ of $B$ with eigenvalue $\lambda_j$, then 
the outcome $x$ is likely to be close
to $\lambda_j$ in the sense that
\[
\mathrm{Pr}(|\lambda_j- x| < 2\pi \eta )>1-\theta\,.
\]
This follows from ineq.~(\ref{thetaeta}) and the fact that the values $a/2^p$ have to be rescaled by $2\pi$.
We conclude
\begin{equation}\label{ErrTerms}
| 
E_{|\psi_j\rangle} (f(X))-f(\lambda_j)| \leq 2\theta \|f\|_\infty +2\pi K \eta\,.
\end{equation}
The second term on the right of ineq.~(\ref{ErrTerms}) corresponds to the case
that $x$ deviates from $\lambda_j$ by at most $2\pi \eta$
(note that the probability for this event is not decreased
by replacing all outcomes lying outside from $I$ with the
closest values in $I$). 
Then the Lipschitz condition ensures that the error of $f(x)$ is small.
If the error of $x$ is large (which happens with probability at most  
$\theta$), the error of $f(x)$ is still upper 
bounded by $2\,\|f\|_\infty$. 
If we replace the eigenvectors $|\psi_j\rangle$ with some vector
$|\psi\rangle=\sum_j c_j |\psi_j\rangle$, the distribution of
 outcomes in the phase estimation procedure is a mixture of
the distributions for each $|\psi_j\rangle$ with weights
$|c_j|^2$. This is easily checked by analyzing the standard 
 phase estimation.
Thus we obtain 
\begin{equation}\label{ErrTermConvex}
| 
E_{|\psi\rangle} (f(X)) -\langle \psi|f(B)|\psi\rangle 
| \leq 2 \theta \|f\|_\infty +2\pi K \eta\,.
\end{equation} 
from ineq.~(\ref{ErrTerms}) by convexity arguments. 

Now we take into account that we have only an approximation $U$
of $V$.  Since the procedure uses $2^{p+1}-1$ copies of 
the controlled-$U$ respective, controlled-$V$, the norm distance 
between the phase estimation circuits is less than $2^{p+1} \delta$. 
Let $q(a)$ and $\tilde{q}(a)$ denote the probability of
obtaining the result $a$ when the phase estimation is implemented
with $U$ and $V$, respectively. Then the $l^1$-distance between
the measures $q$ and $\tilde{q}$ satisfies
(for technical details see \cite{DiagonalEntry})
\[
\|q-\tilde{q}\|_1:=\sum_a |q(a)-\tilde{q}(a)|\leq 2^{p+2}\,\delta\,.
\]
The error in the expectation value of $f(X)$ 
caused by applying $f$ to outcomes $x$  computed from $a$
when  $a$ is sampled according to $\tilde{q}$ instead of $q$ 
is hence less than $2^{p+2}\,\delta \,\|f\|_\infty$. 

Putting everything together we obtain the desired bound
\begin{equation}\label{eq:expectationClose}
|E_{|\psi\rangle} (f(X))-\langle \psi|f(B)|\psi\rangle |
< 
(2\theta+ 2^{p+2}\,\delta) \,  \|f\|_\infty +2\pi K \eta\,.
\end{equation}

\medskip
Since we have shown that the expectation value of the values $f(x)$ 
in the procedure  in Lemma~\ref{FuncLemm} is close to the expectation value
$\langle \psi |f(B)|\psi \rangle$ we can use the average over the values $f(x)$
after a small number of runs as a good estimate for 
the desired expectation value. 
The following lemma states that we obtain an efficient 
procedure for giving an estimation up to any desired
accuracy which is inverse polynomial in $n$.

\begin{Lemma}[Estimation of Expectation Values]\label{Est}${}$\\
Given an efficient algorithm to simulate $\exp(iB)$
on $n$ qubits  
with $\|B\|\leq 1$ in the sense
that the resources to obtain a circuit $U$ with $\|U-\exp(iB)\|\leq \delta$ 
scale polynomially in $n$ and $1/\delta$.  
Given a state $|\psi\rangle$ whose decomposition into $B$-eigenvectors
contains only those with eigenvalues in the interval $I$. 
Let $f$ be as in Lemma~\ref{FuncLemm}.  Then we can estimate
$\langle \psi|f(B)|\psi\rangle$ up to an error $\epsilon\cdot (\|f\|_\infty+K)$  
with probability at least $1-\alpha$ 
such that the time and space resources are bounded by a polynomial in $n$, $1/\epsilon$, and $\log(1/\alpha)$. 
\end{Lemma}

\medskip
\noindent
Proof: The empirical average of the values $f(x)$ obtained after a few runs 
converges exponentially fast to the expectation value $E_{|\psi\rangle}(f(X))$.
Using Hoeffding's bound \cite{Hoeffding} and the fact that $f(X)$ is a random variable
with bounded range (for details cp. also \cite{DiagonalEntry}) 
one can easily show that the required number of runs 
for estimating then expectation value up to an accuracy $\epsilon$ 
scales inverse polynomial in $\epsilon$ and 
polylogarithmically in $1/\alpha$  
when this error should be guaranteed with probability $1-\alpha$. 
 
It remains to bound the resources required to ensure that 
\[
|E_{|\psi\rangle}(f(X)) - \langle \psi |f(B)|\psi\rangle| < \epsilon\cdot 
(\|f\|_\infty+K)\,.
\]
Taking into account that the number of required ancilla qubits $p$ is only 
logarithmic in $1/\theta$ and $1/\eta$ (see eq.~(\ref{pDef})),
ineq.~(\ref{eq:expectationClose}) shows that the error is bounded 
by $(\|f\|_\infty +K)$ times a polynomial  
in $1/\theta$, $1/\eta$, $1/\delta$ and $n$.

In order to make the overall error smaller than $\epsilon\cdot (|f(I)|+K)$ we 
ensure that each of the three terms $2\theta \,\|f\|_\infty$, 
$2^{p+1}\,\delta \,\|f\|_\infty$, and $2\pi K \eta$
on the rhs. of ineq.~(\ref{eq:expectationClose}) is at 
most $\epsilon\cdot(|f(I)|+K)/3$. 

To this end, we choose $\eta =\epsilon/(6\pi)$ and $\theta := \epsilon/6$. 
 Putting these values into 
eq.~(\ref{pDef}) we obtain the number of required ancilla qubits $p$.  We choose $\delta$ such
that $\delta \,2^{p+2}\leq \epsilon/3$.  Note that $2^p$ is polynomial in $1/\epsilon$, so that
$1/\delta$ is polynomial in $1/\epsilon$ and $n$.  The number of controlled-$U$ circuits is a multiple of
$2^p$ and, thus, only polynomial in $1/\epsilon$.   

\medskip
Lemma~\ref{Est} shows that  there is an efficient quantum algorithm for measuring functions of observables provided that the demanded accuracy is not too high.  In the following sections we introduce several problems and prove that they are BQP-complete.  The proofs that they are in BQP rely on the quantum algorithm for measuring functions of observables.

\section{Diagonal Entry Estimation}\label{Sec:DiagonalEntry}

In this section we extend the results of \cite{DiagonalEntry}
saying that the estimation 
of diagonal entries of $A^m$ for a sparse matrix with entries $\pm 1, 0$ 
 is BQP-complete. Here we argue that the construction in \cite{DiagonalEntry}
shows also that the problem remains BQP-hard if when further  restricting to
matrices with only $4$ non-zero entries. This is made precise in 
Lemma~\ref{4entries}.

Furthermore, Section~\ref{Lipschitz}
makes it possible to provide a unifying picture of the results in 
\cite{DiagonalEntry} and our results presented in Sections~\ref{Paths} 
and \ref{Mixing}
since for all three problems we use 
spectral measures of 
functions of observables to show that they are in BQP.

We first state the formal definition of the decision problem related to the estimation of the diagonal entries of powers of sparse matrices.
\begin{Definition}[Diagonal Entry Estimation]\label{DDE}${}$\\
Given a sparse symmetric $N\times N$-matrix $A$ with real entries, an integer $j\in\{1,\ldots,N\}$, and a positive integer $m=\polylog(N)$, estimate the diagonal entry $(A^m)_{jj}$ in the following sense:

Decide if either 
\[
(A^m)_{jj} \geq g + \epsilon\, b^m
\]
or 
\[
(A^m)_{jj} \leq g - \epsilon\, b^m\,,
\]
for given $g\in [-b^m,b^m]$ and $\epsilon=1/\polylog(N)$, where $b$ is
an a priori known upper bound on the operator norm of $A$.
\end{Definition}

\noindent
We showed in \cite{DiagonalEntry}:
\begin{Theorem} 
The problem "Diagonal Entry Estimation" is BQP-complete.  This remains true if the matrices $A$ have only $\pm 1,0$ as entries.
\end{Theorem}
We recall the formal definition of the complexity class BQP \cite{KitaevShen}.
\begin{Definition}[The class BQP]\label{BQP}${}$\\
A language $L$ is in BQP if and only if there is a
uniformly generated family of quantum circuits $Y_r$ acting on
$\poly(r)$ qubits that decide if a string ${\bf x}$ of length $r$ is
contained in $L$ in the following sense:
\begin{equation}\label{Schalt}
Y_r|{\bf x},{\bf 0}\rangle = 
\alpha_{{\bf x},0} |0\rangle\otimes |\psi_{{\bf x},0}\rangle +
\alpha_{{\bf x},1} |1\rangle \otimes |\psi_{{\bf x},1}\rangle
\end{equation}
such that
\begin{enumerate}
\item $|\alpha_{{\bf x},1}|^2 \ge 2/3$ if ${\bf x}\in L$ and
\item $|\alpha_{{\bf x},1}|^2 \le 1/3$ if ${\bf x}\not\in L$\,.
\end{enumerate}
Equation~(\ref{Schalt}) has to be read as follows. The input string ${\bf x}$ determines the first $r$ bits. Furthermore,  $l$ additional ancilla bits are initialized to $0$.  After $Y_r$ has been  applied we interpret
the first qubit as the relevant output and the remaining $r+l-1$ output values  are irrelevant.  The size of the ancilla register is polynomial in $r$.
\end{Definition}

To show that ``Diagonal Entry Estimation'' is in BQP 
we simply apply the quantum algorithm for 
measuring functions of the observable $A/b$ 
to the state $|j\rangle$, and the function 
$f:[-1,1]\rightarrow \R$ with $f(x):=x^m$.  Then 
we can use Lemma~\ref{Est}   
to show that we can efficiently estimate $\langle j|A^m|j\rangle$ with 
sufficient accuracy. To see this, we observe that  
the Lipschitz-constant $K$ satisfies $K\leq m$ and we have $\|f\|_\infty=1$.
Hence we can efficiently estimate the diagonal entries of $(A/b)^m$ 
up to $\epsilon\, (m+1)$ for any desired inverse polynomial $\epsilon$. 
With $\epsilon':=\epsilon\,(m+1)$ we have an accuracy $\epsilon'$. 
Thus, we may achieve an accuracy of $\epsilon'\,b^m$ for the diagonal entries
of $A^m$ for any desired inverse polynomial $\epsilon'$. 
By Definition~\ref{DDE}, 
this is sufficient to solve ``Diagonal Entry Estimation''.

The idea of our proof of the BQP-hardness is based on an encoding of the quantum circuit which solves a given BQP-problem into a self-adjoint operator $A$ such that the spectral measure induced by $A$ provides the information on the solution.  To this end, we assume that we are given a quantum circuit $Y_r$ 
which decides whether a string ${\bf x}$ is in the given language $L$ in the sense of Definition~\ref{BQP}.

\begin{figure}
\centerline{
\Qcircuit @C=3em @R=2.8em {
\lstick{|x_1,\dots,x_r\rangle}&    \multigate{1}{Y_r} &\gate{\sigma_z}
&\multigate{1}{Y_r^\dagger} &\qw  \\
\lstick{|0,\dots,0\rangle}&  \ghost{Y_r} &\qw & \ghost{Y^\dagger_r} &\qw
}
}
\vspace{0.5cm}
\caption{\label{Circ}{\small Circuit $U$ constructed from the original circuit $Y_r$. Whenever the answer of the BQP problem is no, the output state of $U$ is close to the input state $|{\bf x},{\bf 0}\rangle \equiv |x_1,\dots,x_n,0,\dots,0\rangle$.  Otherwise, the state $|{\bf x},{\bf 0}\rangle$ is only restored after applying $U$ twice.}}
\end{figure}
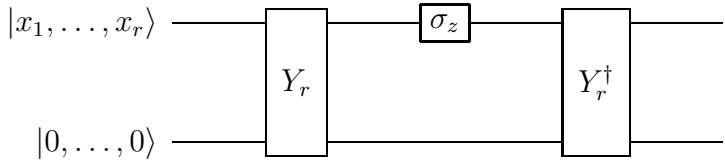

In some analogy to \cite{IdentityQMA,WZ:06} we construct a circuit $U$ which is obtained from $Y_r$ as follows:  first apply the circuit $Y_r$, then apply the $\sigma_z$-gate on the output qubit, and finally
apply the circuit $Y^{\dagger}_r$.  The resulting circuit $U$ is shown in Fig.~\ref{Circ}.  We  denote the dimension of the Hilbert space $U$ acts on by $\tilde{N}$.

Let $U$ be generated by a concatenation of the $M$ elementary gates 
\[
U_0,\dots,U_{M-1}\,.
\] 
We assume furthermore 
that $M$ is odd, which
is automatically satisfied if we decompose 
$Y_r^\dagger$ in analogy to $Y_r$ and
implement a $\sigma_z$-gate between $Y_r$ and $Y_r^\dagger$.  
We define the unitary
\begin{equation}\label{Ul}
W:=\sum_{l=0}^{M-1} |l+1\rangle\langle  l| \otimes U_l\,,
\end{equation}
acting on $\C^M\otimes \C^{\tilde{N}}$. 
Here the $+$ sign in the index has always to be read modulo $M$. 

Now we define the self-adjoint operator
\begin{equation}\label{A}
A:=\frac{1}{2}(W+W^\dagger)\,.
\end{equation}
Then we apply an $A$-measurement to the state
\[
|s_{\bf x}\rangle:=|0\rangle \otimes |{\bf x},{\bf 0}\rangle\,.
\]
Since $|s_{\bf x}\rangle$ is a basis state, we can choose $j$ such that
$|j\rangle=|s_{\bf x}\rangle$. 
To understand why $|\alpha_0|$ influences the measurement statistics it
is useful to consider the extreme cases $|\alpha_1|=0,1$. 
One may check that the repeated application of $W$ to $|j\rangle$ 
leads for $\alpha_0=0$  
to an orbit (of mutually orthogonal states) 
that is periodic after $M$ steps.
The spectral measure induced by $|j\rangle$ and $W$ is hence 
the uniform distribution over the $M$th roots of unity. 
For $|\alpha_1|=1$ the orbit is $M$-periodic up to the phase $-1$.
This leads to a uniform distribution over the roots of unity
reflected at the imaginary axis. 
As shown in \cite{DiagonalEntry} in detail, the  spectral measure induced by $|j\rangle$ and $A$ depends therefore on $\alpha_0,\alpha_1$ 
in the following way:

The spectral measure is the convex sum 
\begin{equation}\label{SpecSol}
P:=|\alpha_0|^2 P^{(0)}+ |\alpha_1|^2 P^{(1)}\,,
\end{equation}
where $P^{(0)}$ is supported by the values $\lambda^{(0)}_l:= \cos(2\pi l/M)$ with probabilities
$P^{(0)}_l=1/M$ for $l=0$ and $P^{(0)}_l=2/M$ for $l=1,\dots,M-1/2$.  The measure $P^{(1)}$ is obtained from $P^{(0)}$ by a reflection at the origin and is hence supported by the values $\lambda_l^{(1)}:= \cos(\pi (2l+1)/M)$ with probabilities $P^{(1)}_{(M-1)/2}=1/M$ and $P^{(1)}_l=2/M$ for $l < (M-1)/2$. The measure $P$ can also be written as
\begin{equation}\label{Pnochmal}
P(\lambda):=\sum_l (|\alpha_0|^2\delta_{\lambda, \lambda_l^{(0)}} P^{(0)}_l
+ |\alpha_1|^2\delta_{\lambda, \lambda_l^{(1)}}P^{(1)}_l)\,.
\end{equation}
Then we  obtain the $j$th diagonal entry of $A^m$ as the $m$th moment of the
measure $P$:

\begin{eqnarray}\nonumber
(A^m)_{jj}&= & \sum_\lambda \lambda^m \,P(\lambda) \\&=&
(1-|\alpha_1|^2)\sum_l \Big(\lambda_l^{(0)}\Big)^m
P_l^{(0)}
+ |\alpha_1|^2\sum_l \Big(\lambda_l^{(1)}\Big)^m
P_l^{(1)}\,.\label{Amjj}
\end{eqnarray}
Since the possible range of  $|\alpha_1|$ is determined by the solution of the decision problem the latter also determines the range of the possible values for $(A^m)_{jj}$. One checks easily that $(A^m)_{jj}$ changes
sufficiently when changing $|\alpha_1|$, i..e,  an estimation of
$(A^m)_{jj}$ makes it possible to decide whether ${\bf x}\in L$. 

To see that the proof of BQP-hardness works also with matrices having only $\pm 1,0$ as entries we make the following observation.  
Hadamard-gates and Toffoli-gates form a universal set. In 
\cite{DiagonalEntry} 
we have instead chosen a universal set containing only Hadamard gates 
and gates which are concatenations of one Hadamard gate with a 
Toffoli gate. Then the above unitaries $U_l$ have only $\pm 1/\sqrt{2}$ as entries. Since the decision 
problem Diagonal Entry Estimation 
is formulated in a scale invariant way, we may rescale $A$ by $\sqrt{2}$
and obtain only entries $\pm 1,0$. 

In the following we will need
the following extension of the results
in \cite{DiagonalEntry}:

\begin{Lemma}\label{4entries} 
The problem Diagonal Entry Estimation remains BQP-hard if 
$A$ contains only entries $\pm 1, 0$ {\em and}
the number of non-zero entries  in each row of $A$ is $4$.
\end{Lemma}

The proof is obvious after observing that
$A$ as constructed above has exactly $4$ nonzero entries in each column and, similarly, in each row.  Each gate $U_l$ has exactly two non-zero entries since it is the embedding of either a Hadamard gate or a Hadamard gate combined with a Toffoli gate.  One checks easily that $W$ has also two non-zero entries in each row leading to four entries in $W+W^\dagger$.

\section{Difference of Numbers of Paths}\label{Paths}

In this section we show that a modified version of Diagonal Entry Estimation
is still BQP-complete when restricting to matrix with entries $1,0$.
This makes the problem more combinatorial than the original one.
In other words,
we introduce a problem concerning regular sparse graphs and show that it is BQP-complete.  A graph $G$ is called regular if the degrees of all its vertices are equal.  We call $G$ sparse if its adjacency matrix 
$\tilde{A}$ is sparse, i.e., for each vertex the number of adjacent vertices
is polylogarithmic in the number of nodes and there is an efficiently computable function that computes the set of adjacent vertices for each vertex. 
We define the following problem:

\begin{Definition}[Difference of Numbers of Paths]\label{NoP}${}$\\
Let $\tilde{A}$  be the adjacency matrix of 
a regular sparse graph $G$ on $N$ vertices and $m$ a positive integer $m=\polylog(N)$.
Let $q$ and $r$ be two vertices such that there exists an automorphism of $G$ which exchanges $q$ and $r$.  Set
\[
\Delta_{qr}^{(m)}:=
(\tilde{A}^m)_{qq} - (\tilde{A}^m)_{qr}\,.
\] 
Decide if either 
\[
\Delta^{(m)}_{qr} \geq g + \epsilon\, b^m
\]
or 
\[ 
\Delta^{(m)}_{qr} \leq g - \epsilon\, b^m\,,
\]
for given $g\in [-b^m,b^m]$ and $\epsilon=1/\polylog(N)$, where $b$ is
an a priori known upper bound on 
\[
\sup_n |\Delta^{(n)}_{qr}|^{1/n}\,.
\]
\end{Definition}

\begin{Theorem} The problem ``Difference of Numbers of Paths'' is BQP-complete.
\end{Theorem}
We prove this theorem by showing that this problem is BQP-hard and is contained in BQP.  The proof of BQP-hardness is based on a reduction to the problem ``Diagonal Entry Estimation''.  An efficient quantum algorithm for estimating the difference of number of paths is obtained by applying the algorithm for measuring functions of observables.

\subsection{``Difference of Numbers of Paths'' is BQP-hard}
We show that ``Diagonal Entry Estimation'' reduces to ``Difference of Number of Paths''.  Let $A$ be an $N\times N$ matrix with entries $\pm 1,0$ as in Definition~\ref{DDE}.  The reduction relies on the conversion of $A$ into a $0$-$1$-matrix $\tilde{A}$ acting on a space of dimension $D:=2N$ such that the restriction of $\tilde{A}$ to some invariant $N$-dimensional subspace is unitarily equivalent to $A$. 

The conversion works as follows.  Each $-1$ in $A$ is replaced by the Pauli-matrix $\sigma_x$ and
each $1$ by the identity matrix ${\bf 1}_2=\sigma_x^2$.  Each $0$ is replaced by the zero matrix ${\bf 0}_2$.  More formally, 
the $\pm 1$-$0$-matrix
\[
A=\sum_{i,j=0}^{N-1} a_{ij} |i\rangle \langle j|
\]
is converted to the $0$-$1$-matrix
\[
\tilde{A}:= \sum_{i,j=0}^{N-1} \gamma_{ij} \otimes |i\rangle \langle j|\,,
\]
where $\gamma_{ij}:=\sigma_x$ if $a_{ij}=-1$, $\gamma_{ij}:={\bf 1}_2$ 
if $a_{ij}=1$, and $\gamma_{ij}:={\bf 0}_2$ if $a_{ij}=0$. 

We now prove that the restriction of $\tilde{A}$  
to some $N$-dimensional subspace is unitarily equivalent to $A$.  To this end, we start by introducing the vectors
\[
|\phi^\pm\rangle:=\frac{1}{\sqrt{2}}(|0\rangle \pm |1\rangle) \in \C^2\,.
\]
On the subspace spanned by $|\phi^-\rangle$ the matrices $\gamma_{ij}$ act by multiplying the vectors by the scalars $a_{ij}$, i.e.,
\[
\gamma_{ij}|\phi^-\rangle=a_{ij} |\phi^-\rangle\,.
\]
Similarly, on the subspace spanned by $|\phi^+\rangle$ they act by multiplying the vectors by the scalars $|a_{ij}|=a_{ij}^2$, i.e.,
\[
\gamma_{ij}|\phi^+\rangle=a_{ij}^2 |\phi^+\rangle\,.
\]
Consequently, we have the direct sum decomposition
\[
\tilde{A}=|\phi^-\rangle\langle \phi^-| \otimes A + |\phi^+\rangle\langle 
\phi^+| 
\otimes (A*A)\,,
\]
where $A*A$ is the Hadamard product (or entry-wise product) 
of $A$ with itself
and has therefore only $0,1$ as entries. 
In the following, it will be important that $\tilde{A}$ is invariant under conjugation with $\sigma_x\otimes{\bf 1}_N$.  This invariance follows directly from the direct sum decomposition and the fact the vectors $|\phi^\pm\rangle$ are eigenvectors of $\sigma_x$ with corresponding eigenvalues $\pm 1$.

Assume we want to estimate $(A^m)_{jj}$. Let $|j\rangle$ 
be the $j$th basis vector in $\C^N$. 
Then we define vectors in $\C^2\otimes \C^N\equiv \C^{2N}$ 
by 
$|\psi^-\rangle:=|\phi^- \rangle \otimes |j\rangle$, $|q\rangle:=|0\rangle \otimes |j\rangle, |r\rangle:=|1\rangle \otimes |j\rangle$, and label the corresponding vertices by $q$ and $r$, respectively.
We have then $|\psi^-\rangle=(|q\rangle +|r\rangle)/\sqrt{2}$.
Note that the automorphism $\sigma_x\otimes {\bf 1}$ exchanges the vertices $q$ and $r$.  
We reproduce diagonal entries of $A^m$ from both diagonal and off-diagonal entries of $\tilde{A}^m$ as follows:
\begin{eqnarray}
(A^m)_{jj}&=&\langle\psi^-| \tilde{A}^m  
|\psi^-\rangle    \nonumber  
\\&=& \sqrt{2}\langle\psi^-| \tilde{A}^m \nonumber 
|q\rangle  \label{-0}
\\&=& (\tilde{A}^m)_{qq} -(\tilde{A}^m)_{qr}\label{Number}=\Delta_{qr}^{(m)}\,.
\end{eqnarray}
The second equality follows from the fact that the space $|\phi^-\rangle \otimes\C^N$ is $\tilde{A}$-invariant.  Thus, only the component of 
$|q\rangle$ lying in this subspace is relevant. 

The terms in eq.~(\ref{Number}) correspond to the number of paths of length $m$ from $q$ to $q$ and from $q$ to $r$, respectively.   
In other words, the problem of estimating diagonal entries of $A^m$ reduces to estimating the number of paths of length $m$ from $q$ to $q$ and $q$ to $r$ in the graph defined by $\tilde{A}$, where $q$ and $r$ have to be chosen suitably.

Taking into account that also  off-diagonal entries of powers
of sparse matrices can be estimated efficiently \cite{DiagonalEntry},
it seems as if we had already shown that estimating entries of powers of adjacency matrices is BQP-complete. 
However, this is not the case.  There is a subtle but very important issue concerning the precision of this estimation.  The accuracy in Definition~\ref{DDE} is given by  $\epsilon\, b^m$, where $b$ is an a priori known bound on the operator norm of $A$.  In general, this bound will not be a bound on the norm of $\tilde{A}$.  Using the direct sum decomposition of $\tilde{A}$, we see that the norm of $\tilde{A}$ is equal to $\max\{\|A\|,\|A*A\|\}$ and is smaller or equal to $\max\{b,b^2\}$ as $\|A*A\|\le \|A\otimes A\|$ \cite{Horn}.  
On the one hand, if we estimate the entries of $\tilde{A}^m$ up 
to an accuracy specified in terms of the norm of $\tilde{A}$, then these values are not good enough to obtain a sufficiently precise estimate of the diagonal entry $(A^m)_{jj}$. Assume for instance, that $c\|A\|=\|\tilde{A}\|$
for some constant $c>1$. Then the entries of $\tilde{A}$ can only be estimated
with an accuracy that is an inverse polynomial multiple of $c^m\|A\|^m$,
which is exponentially worse than the required accuracy $\epsilon\,\|A\|^m$.  
On the other hand, the operator norm $\|A\|$ of $A$ is not a natural quantity to define the required accuracy of the estimation of the number of paths in $\tilde{A}$. 

For these reasons the estimation accuracy of the decision problem ``Difference of Numbers of Paths'' is not (and should not be) formulated in terms of the norm of $\tilde{A}$.  Instead, it is specified in terms of a quantity which is sufficiently small, making it possible to reduce Diagonal Entry Estimation to the former. 
For doing so,
we recall that for each $n\in \N$ the $n$th diagonal entry of $A^n$ 
can be expressed in terms of $\Delta^{(n)}_{qr}$ due to
\[
(A^n)_{jj}=\Delta^{(n)}_{qr}\,.
\]
We have therefore
\[
\sup_{n\in \N}| \Delta^{(n)}_{qr}| =
\sup_{n\in \N} |\langle \psi^-| \tilde{A}^n|\psi^-\rangle|=
\sup_{n\in \N} |(A^n)_{jj}| \leq   \|A\|^n\,.
\]
This proves that if we can solve ``Difference of Number of Paths'' then we can also decide between the two cases in Definition~\ref{DDE} with $b:=\|A\|$. 

It remains to show that $\tilde{A}$ is the adjacency matrix of a {\it regular} graph.  We first recall Lemma~\ref{4entries}.
By replacing the values $\pm 1$ with the $2\times 2$ identity matrix and the matrix $\sigma_x$, respectively, we obtain a matrix with four entries in each row, as well.  Hence all vertices of the graph have degree $4$.

\subsection{``Difference of Number of Paths'' is in BQP}

\label{PathsInBQP}
To solve the problem ``Difference of Numbers of Paths'' on a quantum computer we 
observe that $\Delta_{pr}^{(m)}$ is an expectation value of a
random variable which occurs in an appropriate measurement procedure.
In analogy to the preceding paragraph, we define
\[
|\psi^-\rangle:=\frac{1}{\sqrt{2}}(|q\rangle+|r\rangle)\,,
\]
with the essential difference that the nodes $q$ and $r$ 
are a priori given and not derived 
from  setting $|q\rangle:=|0\rangle \otimes |j\rangle$ and $|r\rangle :=|1\rangle \otimes|j\rangle$. 
Then we have nevertheless 
\begin{eqnarray}
\Delta_{qr}^{(m)}
&=& \langle \psi^-|\tilde{A}^m|\psi^-\rangle \label{Delta}\,,
\end{eqnarray}
which follows here from the symmetry defined by the given graph 
automorphism.  
Without this symmetry, we had
\[
\langle \psi^-|\tilde{A}^m|\psi^-\rangle=
\frac{1}{2}(\Delta_{qq}^{(m)}-2\Delta_{qr}^{(m)}+\Delta_{rr}^{(m)})\,,
\]
but here we have $\Delta_{qq}^{(m)}=\Delta_{rr}^{(m)}$. 
Due to the promise on the growth of $\Delta_{qr}^{(n)}$ the decomposition of $|\psi^-\rangle$ in $\tilde{A}$-eigenvectors contains only those eigenvectors whose corresponding eigenvalues satisfy $\lambda_j \in [-b,b]$.  
We may rescale $\tilde{A}$ 
by some factor $d$ 
 such that its norm is at most $1$. 
On the rescaled interval $[-b/d,b/d]$ we define the function $f(x):=x^m$
having Lipschitz-constant $K\leq m (b/d)^m$ and sup-norm $\|f\|_\infty=(b/d)^m$. Then the achievable estimation error for 
$\langle \psi^- |(\tilde{A}/d)^m|\psi^-\rangle=\Delta_{qr}^{(m)}/d^m$  
is an inverse polynomial
multiple of $(b/d)^m$, i.e., we can estimate $\Delta_{qr}^{(m)}$ up to 
any desired inverse polynomial multiple of $b^m$.

\subsection{Interpretation of difference of number of paths in terms of discrete-time random walks}
Note that there is also an alternative interpretation for the difference of the number of paths.  Since $\tilde{A}$ describes a regular graph of degree $4$, we obtain a doubly stochastic matrix by 
\[
\hat{A}:=\frac{1}{4}\tilde{A}\,.
\]
It describes a classical random walk (in discrete time) 
on the corresponding graph.
Then the entry $(\hat{A}^m)_{ij}$ is the probability of reaching position $j$
after $m$ steps  given that the initial position was $i$. 
For increasing $m$, the difference $(\hat{A}^m)_{ii} -(\hat{A}^m)_{ij}$
is therefore directly linked to the decay of probability differences, i.e.,
to mixing properties of the random walk.

\section{Decay of Probability Differences}\label{Mixing}
We can also formulate a BQP-complete problem in term of 
{\it continuous-time} random walks on regular sparse graphs.  The dynamics of a random walk is generated by the Laplacian of the underlying graph.  Consequently, the time evolution of an initial probability distribution is determined by its spectral resolution into the eigenvectors of the Laplacian.  

For regular graphs, the Laplacian coincides with the adjacency matrix up to the negative sign and an additive multiple of the identity matrix.  Hence, the spectral resolution with respect to the adjacency matrix (being 
crucial in the preceding section) also determines the behavior of the continuous random walk.  We introduce some notation to make these statements more precise.

Let $\tilde{A}$ describe a regular graph on $N$ vertices 
with degree $d$. We define its Laplacian by \cite{Chung}
\[
\cL:=d\, {\bf 1} -\tilde{A}\,.
\]
Let $p(t)=(p_1(t),\ldots,p_N(t))$ be a probability distribution where $p_j(t)$ for $j=1,\dots,N$ is the probability of being at vertex $j$ for $j=1,\ldots,N$.  The Laplacian $\cL$ defines a continuous-time (classical) random walk by
\[
p(t):=e^{-\cL t}p\,,
\]
where $p=p(0)$ is the initial probability distribution.

Our BQP-complete problem considers the following question. 
Assume that the random walk starts at vertex $q$.  Given a second vertex $r$, determine the decay of the difference between the probabilities $p_q(t)$ and $p_r(t)$.

For infinite time, only the smallest eigenvalue whose eigenvector is contained in the initial state $p$ determines the exponent dominating the decay.  For our specific problem (where only the difference between the probabilities of two vertices is considered), only the spectral decomposition of $|q\rangle -|r\rangle$ is relevant.  For a random walk which started at vertex $q$, the difference between $p_q(t)$ and $p_r(t)$ is given by 
\begin{equation}\label{Lp}
c_{qr} (t):=(e^{-\cL t})_{qq}-(e^{-\cL t})_{qr}\,,
\end{equation}
where we use $(W)_{qr}$ to denote the entry of an arbitrary matrix $W$
in row $q$ and column $r$. 

Our BQP-complete problem is defined as follows.
\begin{Definition}[Decay of Probability Differences]\label{Def:Mix}${}$\\
Let $c_{qr}(t)$ be defined as in eq.~(\ref{Lp}), where $\cL$ is the Laplacian 
of a regular graph $G$ on $N$ vertices with degree $d\in O(\polylog (N))$
such that there is an efficiently computable function that
determines the set of adjacent vertices for every
vertex.  
  Let $\mu>0$ and $b<a<1$ be  two positive numbers 
with $1/a,1/b,1/(a-b) \in \polylog(N)$ and $T\in \polylog(N)$ be some time instant.  Let $q,r$ be two vertices for which there is an automorphism of $G$ exchanging $q$ and $r$.  Given the promise that 
\[
c_{qr}(t) \in O(e^{-\mu t})\,,
\]
decide if either
\[
c_{qr}(T) \geq a\,e^{-\mu T} 
\]
or
\[
c_{qr} (T) \leq b\,e^{-\mu T}  \,.
\] 
\end{Definition}
Note that in both cases, the exponent of decay may be the same but only the 
constants $a$ and $b$ differ. 

\begin{Theorem} 
The problem ``Decay of Probability Differences'' is BQP-complete.
\end{Theorem}
We prove this theorem by showing that the problem  
is BQP-hard and is contained in BQP.

\subsection{``Decay of Probability Differences'' is BQP-hard}
To see that every problem in BQP can be reduced to ``Decay of Probability Differences''
we consider the adjacency matrix $\tilde{A}$ obtained from the matrix $A$ as described in Section~\ref{Paths}.  Adopting the notation used 
there, we have 
 $|q\rangle:=|0\rangle\otimes |j\rangle$ and
$|r\rangle:=|1\rangle \otimes |j\rangle$ and label the corresponding vertices by $q$ and $r$, respectively
and obtain in straightforward analogy the following equality:
\begin{eqnarray}
c_{qr}(t)&=& \nonumber \langle q| e^{-\cL t}|q\rangle - \langle q| e^{-\cL t}|r\rangle\\
&=& \nonumber \sqrt{2}\langle \psi^-| e^{-\cL t}|q\rangle \\
&=& \nonumber \langle \psi^-| e^{-\cL t} |\psi^-\rangle \\
&=& \nonumber \langle \psi^-| e^{-(4 \cdot {\bf 1}_{2N}-\tilde{A})t} |\psi^-\rangle \\
&=& \label{DecEq}\langle j| e^{-(4 \cdot {\bf 1}_{N}-A)t} |j\rangle\,.
\end{eqnarray}
We need some properties of the spectral measure induced by $A$ and $|j\rangle$ to prove that ``Decay of Probability Differences'' is BQP-hard. 
They are obtained from Section~\ref{Sec:DiagonalEntry}
after taking into account the rescaling factor $\sqrt{2}$ 
(see the end of the section)
 and read:
\begin{enumerate}
\item the largest eigenvalue in the support of the spectral measure is $\sqrt{2}$ and the second largest is $\sqrt{2}\cos(\pi/M)$, and   
\item the probability of obtaining the eigenvalue $\sqrt{2}$ is equal to $|\alpha_0|^2/M$.
\end{enumerate}
Set $\mu:=4-\sqrt{2}$ and $\nu:=4- \sqrt{2} \cos(\pi/M)$ which are equal to the smallest and second smallest eigenvalue, respectively, that occur
in the spectral measure  induced by $4\cdot {\bf 1}_N -A$ 
and $|j\rangle$. 
Using these eigenvalues, eqs.~(\ref{DecEq}) imply
\[
c_{qr}(t) \geq \frac{|\alpha_0|^2}{M} e^{-\mu t}
\]
and 
\[
c_{qr}(t) \leq \frac{|\alpha_0|^2}{M} e^{-\mu t} + e^{-\nu t}\,.
\]
Choose the time instant $T:=\ln (6 M) /(\nu-\mu)$.  In this case, 
the bounds are
\[
c_{qr}(T) \geq  \frac{|\alpha_0|^2}{M} e^{-\mu T}
\]
and 
\[
c_{qr}(T) \leq  \frac{|\alpha_0|^2+1/6}{M} e^{-\mu T} \,.
\]
Note that $T$ increases only polynomially with $M$ since
$\nu-\mu$ scales inverse polynomially with $M$. 
 Taking into account that
$|\alpha_0|^2 \leq 1/3$ if $x\in L$ and $|\alpha_0|^2 \geq 2/3$ otherwise,
we have for ${\bf x}\in L$
\[
c_{qr}(T) \leq \frac{1}{2M} e^{-\mu T} \,,
\]    
and for ${\bf x} \not \in L$ 
\[
c_{qr}(T) \geq  \frac{2}{3M} e^{-\mu T} \,.
\]    
This proves that we can reduce the question whether ${\bf x} \in L$ to
the estimation of $c_{qr}(T)$.

\subsection{``Decay of Probability Differences'' is in BQP}
In strong analogy to Section~\ref{Paths}, 
the quantity of interest, which is here $c_{qr}(T)$, can be written
as an expectation value
\begin{eqnarray}\label{Lt}
c_{qr}(T)
&=&
\langle q | e^{-\cL T} | q\rangle -
\langle q | e^{-\cL T} | r\rangle  \nonumber \\
&=& 
\langle \psi^- | e^{-\cL T} |\psi^-\rangle\,,
\end{eqnarray}
where the last equation 
follows (in analogy to 
Subsection~\ref{PathsInBQP}) from the symmetry of the graph and not, 
as in the preceding paragraph, 
from setting $|q\rangle:=|0\rangle \otimes |j\rangle$. 
The last expression in eq.~(\ref{Lt}) is an expectation value of a random experiment
where the observable $\cL$ is measured and the result $\lambda$ is
converted to $\exp(-\lambda t)$.  
To show that ``Decay of Probability Differences'' is in BQP we 
rescale $\cL$ to $\cL /b$ where $b$ is a polynomial upper bound of the
norm of $\cL$. We have, for instance,  $\|\cL\|\leq 2d$ since the norm of
$\tilde{A}$ is given by its degree $d$ \cite{CDS}. 
From the promise saying that the  probability differences decay with
$O(\exp(-\mu t))$  
it follows that the decomposition of
$|\psi^-\rangle$ contains only $\cL$-eigenvectors with eigenvalues 
in $[\mu,\infty]$, i.e., we have only eigenvalues of
$\cL /b$ in $[\mu/b,\infty]$. 
We define $f:[\mu/b,\infty]\rightarrow \R^+$ 
with $f(x):=\exp(-x\, b T)$. Its Lipschitz constant is 
$K= bT \exp(-\mu T)$. Moreover, we have  $\|f\|_\infty=\exp(-\mu T)$. 
Lemma~\ref{Est} guarantees that we can efficiently estimate 
the expectation value of $f(\cL /b)$ up to any desired inverse
polynomial multiple of $(bT+1)\exp(-\mu T)$. Since $b$ and $T$ are both 
polynomial, we can obtain an accuracy being any desired 
inverse polynomial multiple of 
$\exp(-\mu T)$. This is sufficient to
solve ``Decay of Probability Differences''.

\section{A QCMA-complete mixing problem}
\label{Sec:QCMA}
The complexity class QCMA, which is one possible quantum analogue of the
classical class NP \cite{AN:02} can be obtained by modifying
Definition~\ref{BQP} ``slightly''.
Roughly speaking, we change the problem by  not asking 
whether a {\it given state} 
$|{\bf x},{\bf 0}\rangle$ is accepted by the circuit $Y_r$ 
in the sense that the
output qubit is with high probability 
$|1\rangle$ after applying $Y_r$ to the state. 
Instead, the problem is 
whether there {\it exists} a basis 
state
 that is accepted. 
In this case, this basis 
state is called a ``witness''
for the decision problem. 
The input string ${\bf x}$ 
does not define the {\it input state} for the circuit. Instead,  
it defines the circuit $Y_{\bf x}$ itself. 
More precisely, we have:

\begin{Definition}[Complexity Class QCMA]\label{QCMA}${}$\\
A language $L$ is in QCMA if there is a uniformly generated
family $(Y_{\bf x})_{\bf x}$ of circuits (where ${\bf x}$ 
runs over the problem instances) 
acting on $r+l\in O(\poly(|x|))$  qubits
such
that the following
statements are true:

If we define for every $r+l$-qubit basis state $|{\bf y},{\bf 0}\rangle$
the 
$r+l-1$-qubit
states $|\psi_{{\bf x},{\bf y},0}\rangle$ and
$|\psi_{{\bf x},{\bf y},1}\rangle$  by
\[
Y_{\bf x} |{\bf y},{\bf 0}\rangle =\alpha_{{\bf x},{\bf y},0} 
|0\rangle \otimes |\psi_{{\bf x},{\bf y},0}\rangle 
+ \alpha_{{\bf x},{\bf y},1}   |1\rangle \otimes |\psi_{{\bf x},{\bf y},1}\rangle\,, 
\]
then
we have:

\begin{enumerate}

\item
For every ${\bf x}\in L$ 
there is a ``witness'' ${\bf y}$ such that   
\[
|\alpha_{{\bf x},{\bf y},1}|\geq 2/3\,.
\]

\item 
For every  ${\bf x}\not\in L$ 
one has
\[
|\alpha_{{\bf x},{\bf y},1}|\leq 1/3
\]
for
every ${\bf y} \in \{0,1\}^r$.

\end{enumerate} 
\end{Definition}
Then we find:

\begin{Theorem}[QCMA-Complete Mixing Problem]\label{Th:QCMA}${}$\\
The following problem is QCMA-complete:

Given a regular sparse graph $G$ on $2N$ nodes such that there exists
an automorphism of $G$ that exchanges all nodes $2j$ with $2j+1$
for $j=0,1,\dots,N-1$. 
Let $\mu>0$ and $b<a<1$ be  two positive numbers 
with $1/a,1/b,1/(a-b) \in \polylog(N)$ and $T\in \polylog(N)$ be some time instant.

Moreover, let $\tilde{N}\leq N$ be given with the
 promise that for all
$j\leq \tilde{N}$ we have $c_{2j,2j+1}(t)\in O(\exp(-\mu t))$. 
Decide if either
\begin{enumerate}

\item there is a number $j\leq \tilde{N}$ with
\[
c_{2j,2j+1}(T) \geq a \,e^{-\mu T}
\]
or
\item
for all $j\leq \tilde{N}$ we have
\[
c_{2j,2j+1}(T) \leq b \,e^{-\mu T}\,.
\]
\end{enumerate}
\end{Theorem}

To show that the problem is QCMA-hard we
construct  the graph $G_{\bf x}$ corresponding to a given $Y_{\bf x}$ 
 according to
Section~\ref{Paths}, but keep in mind that the 
input string {\it for the circuit}  (i.e., the potential witness) 
is here called 
$|{\bf y},{\bf 0}\rangle$ 
instead of $|{\bf x},{\bf 0}\rangle$.
We 
reorder the nodes such that  the states
$|2j\rangle$ correspond to the states 
$|0\rangle \otimes |{\bf y},{\bf 0}\rangle$ 
and
the states $|2j+1\rangle$ to $|1\rangle \otimes |{\bf y},{\bf 0}\rangle$
for $j=0,1,\dots,\tilde{N}$, where $\tilde{N}$ is chosen appropriately. 
Section~\ref{Mixing} describes in detail how (1.) and (2.) in the above theorem correspond
to the cases that the state $|{\bf y},{\bf 0}\rangle$ 
is accepted or not by the circuit. Hence we have for each ${\bf x}$ 
a graph such that a pair $(2j,2j+1)$ of nodes exists with $j\leq \tilde{N}$ 
and slow decay of probability differences if and only
if there exists a witness ${\bf y}$ for the QCMA-problem.

To see that the problem is in QCMA we recall from the preceding section 
that we  can efficiently determine 
$c_{2j,2j+1}(t)$ on the appropriate scale by estimating appropriate
expectation values  when applying the phase estimation procedure
to the corresponding states 
\begin{equation}\label{2j}
\frac{1}{\sqrt{2}}(|2j\rangle - |2j+1\rangle)\,.
\end{equation}
The numbers $j$ are therefore the potential witnesss:
${\bf x}$ is in $L$ if and only if there exists a
$j$  which leads to slow decay of probability differences. 
We have now to construct a circuit where the values $j$ define
the possible input basis states.

We define a quantum register with $\lceil \log \tilde{N} \rceil +1$ qubits 
where we identify each state (\ref{2j})  with 
\[
\frac{1}{\sqrt{2}}(|0\rangle - |1\rangle)\otimes |j\rangle\,.
\] 
Assume we have already constructed a
quantum circuit 
on $\lceil \log \tilde{N} \rceil+1+r$ qubits (with some appropriate $r$) 
describing the whole phase estimation procedure
including the post-processing of the outcomes with the function $f$
and the repeated sampling. This construction is rather technical,
but in principle straightforward. Now we modify the circuit such
that it accepts only those $j$  with $j\leq \tilde{N}$. Finally,
we add a Hadamard gate acting on the left component such that 
the state can be  initialized to $|0\rangle$ instead of the superposition
$|0\rangle -|1\rangle$. 
We define the correspondence 
\[
|0\rangle \otimes |j\rangle \otimes |0\cdots 0\rangle \equiv
|{\bf y},{\bf 0}\rangle \,,
\]
where $j$ is now allowed to run over all values $0,1,\dots,2^{\lceil \log \tilde{N}\rceil}$ instead of only $0,1,\dots,\tilde{N}$. 
We have now obtained a circuit for which
there exists 
a witness ${\bf y}$ (defining $r$ bits of the $r+l$-qubit basis 
state as input for 
$Y_{\bf x}$) 
iff there exists
a pair $(2j,2j+1)$ with $j\leq \tilde{N}$ 
that leads to slow decay.

\section{Conclusions}
We have constructed BQP-complete problems concerning the mixing 
properties of classical random walks. Roughly speaking, the problems 
are to estimate  how 
fast the difference between the probability of being at different nodes
decays in discrete and in  continuous-time random walks. 
Given that the quantum computer is more powerful than the 
classical computer in the sense that $BPP\neq BQP$ 
we have hence shown that the quantum computer is also more powerful
in analyzing certain mixing properties
 of classical random walks.



\begin{thebibliography}{10}

\bibitem{KnillQuadr}
E.~Knill and R.~Laflamme.
\newblock Quantum computation and quadratically signed weight enumerators.
\newblock {\em Inf. Process. Lett.}, 79(4):173--179, 2001.

\bibitem{PawelYard}
P.~Wocjan and J.~Yard.
\newblock The Jones polynomial: quantum algorithms and applications in quantum
  complexity theory.
\newblock {\em quant-ph/0603069}.

\bibitem{aharonov-2006-}
Dorit Aharonov and Itai Arad.
\newblock {The BQP-hardness of approximating the Jones polynomial}, 2006.
\newblock
  {\em quant-ph/0605181}.

\bibitem{WZ:06}
P.~Wocjan and S.~Zhang.
\newblock {Several natural BQP-complete problems}.
\newblock {\em quant-ph/0606179}.

\bibitem{DiagonalEntry}
D.~Janzing and P.~Wocjan.
\newblock Estimating diagonal entries of powers of sparse matrices is
  bqp-complete.
\newblock {\em quant-ph/0606229}.

\bibitem{ATS}
D.~Aharonov and A.~Ta-Shma.
\newblock Adiabatic quantum state generation.
\newblock In {\em Proceedings 35th Annual ACM Symp. on Theory of Computing},
  pages 20--23, 2003.

\bibitem{ChildsDiss}
A.~Childs.
\newblock {\em Quantum information processing in continuous time}.
\newblock PhD thesis, Massachusetts Institute of Technology, 2004.

\bibitem{BACS:06}
D.~W. Berry, G.~Ahokas, R.~Cleve, and B.~C. Sanders.
\newblock {Efficient quantum algorithms for simulating sparse Hamiltonians}.
\newblock {\em quant-ph/0508139}.

\bibitem{Schoelkopf}
{B.~Sch\"{o}lkopf and A.~Smola}.
\newblock {\em Learning with kernels}.
\newblock MIT Press, Cambridge, MA, 2002.

\bibitem{ClevePhase}
R.~Cleve, A.~Ekert, C.~Macchiavello, and M.~Mosca.
\newblock Quantum algorithms revisited.
\newblock {\em Proc. Roy. Soc. London A}, 454:339--354, 1998.
\newblock see also {\em quant-ph/9708016}.

\bibitem{NC}
M.~Nielsen and I.~Chuang.
\newblock {\em Quantum Computation and Quantum Information}.
\newblock Cambridge University Press, 2000.


\bibitem{Hoeffding}
W.~Hoeffding.
\newblock Probability inequalities for sums of bounded random variables.
\newblock {\em Journ. Am. Stat. Ass.}, 58(301):13--30, 1963.


\bibitem{KitaevShen}
A.~Kitaev, A.~Shen, and M.~Vyalyi.
\newblock {\em Classical and Quantum Computation}, volume~47.
\newblock Am. Math. Soc., Providence, Rhode Island, 2002.


\bibitem{IdentityQMA}
D.~Janzing, P.~Wocjan, and T.~Beth.
\newblock {``Non-Identity check'' is QMA-complete}.
\newblock {\em Int. Journ. Quant. Inf.}, 3(3):463--473, 2005.

\bibitem{Horn}
R.~Horn and C.~Johnson.
\newblock {\em Topics in Matrix Analysis}.
\newblock Cambridge University Press, 1991.

\bibitem{Chung}
F.~Chung.
\newblock {\em Spectral graph theory}.
\newblock Number~92 in CBMS Regional conference series in Math. Am. Math. Soc.,
  Providence, Rhode Island, 1997.

\bibitem{CDS}
Cvetkovic D., M.~Doob, and H.~Sachs.
\newblock {\em Spectra of Graphs}.
\newblock Johann Ambrosius Barth Verlag, 3rd edition, 1995.

\bibitem{AN:02}
D.~Aharonov and T.~Naveh.
\newblock {Quantum NP - A Survey}.
\newblock {\em quant-ph/0210077}.

\end{thebibliography}

\end{document}